\documentclass[
    ,final            
  ]
  {aipproc}

\layoutstyle{8x11single}

\usepackage{placeins}

\begin{document}
\newcommand{\gevc}{(GeV/c)$^2$~}
\newcommand{\gevcp}{(GeV/c)$^2$}
\newcommand{\abs}[1]{\vert#1\vert^2}
\newcommand{\f}[1]{F_{#1}}
\newcommand{\fb}[1]{\overline{F_{#1}}}
\newcommand{\conj}[1]{#1^*}
\newcommand{\mlp}{M_{1+}}
\newcommand{\elp}{E_{1+}}
\newcommand{\llp}{L_{1+}}
\newcommand{\mlm}{M_{1-}}
\newcommand{\llm}{L_{1-}}
\newcommand{\eOp}{E_{0+}}
\newcommand{\lOp}{L_{0+}}

\title{Accuracy of Extracted Multipoles from
  $\gamma^*N\rightarrow\Delta$ Data}


\newcommand{\mitlns}{Department of Physics, Laboratory
for Nuclear Science, Massachusetts Institute of Technology, Cambridge,
Massachusetts 02139, USA}

\newcommand{\riken}{Current address: Radiation Laboratory, RIKEN, 2-1 Hirosawa, Wako, Saitama 351-0198, Japan}
\newcommand{\uk}{Department of Physics and Astronomy, University of
  Kentucky, Lexington, Kentucky 40206 USA}

\newcommand{\duke}{Current address: Triangle Universities Nuclear Laboratory/Duke
  University, Durham, North Carolina 27708, USA}

\author{S. Stave}{
  address={\mitlns},altaddress={\duke}
}

\author{A. M. Bernstein}{
  address={\mitlns}
}

\author{I. Nakagawa}{
  address={\mitlns},
  address={\uk},
  ,altaddress={\riken}
}

\begin{abstract}
This work evaluates the model dependence of the electric and Coulomb
quadrupole amplitudes (E2, C2) in the predominantly M1 (magnetic
dipole-quark spin flip) $\gamma^* N \rightarrow \Delta$
transition. Both the model-to-model dependence and the intrinsic model
uncertainties are evaluated and found to be comparable to each other
and no larger than the experimental errors. It is confirmed that the quadrupole amplitudes have been accurately measured indicating significant non-zero angular momentum components in the proton and $\Delta$.  

\end{abstract}

\maketitle


\section{Physics Motivation}

Experimental confirmation of the presence of non-spherical hadron
amplitudes (i.e. d states in quark models or p wave $\pi$-N states) is
fundamental and has been the subject of intense experimental and
theoretical interest (for reviews see \cite{nstar2001,cnp,amb}). This
effort has focused on the measurement of the electric and Coulomb
quadrupole amplitudes (E2, C2) in the predominantly M1 (magnetic
dipole-quark spin flip) $\gamma^* N \rightarrow \Delta$ transition.
Since the proton has spin 1/2, no quadrupole moment can be measured.
However, the $\Delta$ has spin 3/2 so the  $\gamma^*N\rightarrow
\Delta$ reaction can be studied for quadrupole amplitudes in the
nucleon and $\Delta$.   Due to spin and parity conservation in the
$\gamma^*N(J^\pi=1/2^+) \rightarrow \Delta(J^\pi=3/2^+)$ reaction,
only three multipoles can contribute to the transition: the magnetic
dipole ($M1$), the electric quadrupole ($E2$), and the Coulomb
quadrupole ($C2$) photon absorption multipoles.  The corresponding
resonant pion production multipoles are  $M_{1+}^{3/2}$,
$E_{1+}^{3/2}$, and $S_{1+}^{3/2}$.  The relative quadrupole to dipole
ratios are EMR=Re($E_{1+}^{3/2}/M_{1+}^{3/2}$) and
CMR=Re($S_{1+}^{3/2}/M_{1+}^{3/2}$).  In the quark model, the
non-spherical amplitudes in the  nucleon and $\Delta$ are caused by
the non-central, tensor interaction between quarks \cite{glashow}.
However, the magnitudes of this effect for the predicted E2 and C2
amplitudes\cite{capstick_karl} are at least an order of magnitude too
small to explain the experimental results 
and even the dominant M1 matrix element is $\simeq$ 30\% low
\cite{amb,capstick_karl}. A likely cause of these dynamical
shortcomings is that the quark model does not respect chiral symmetry,
whose spontaneous breaking leads to strong emission of virtual pions
(Nambu-Goldstone Bosons)\cite{amb}. These couple to  nucleons as
$\vec{\sigma}\cdot \vec{p}$ where  $\vec{\sigma}$ is the nucleon spin,
and $\vec{p}$ is the pion momentum. The coupling is strong in the p
wave and mixes in non-zero angular momentum components.

However, the multipoles are not observables and must be extracted from the
measured cross sections.  The five-fold differential cross section for the
$p(\vec{e},e'p)\pi^{0}$ reaction is written as five two-fold
differential cross sections with an explicit $\phi^*$
dependence as \cite{drechsel_tiator}

\begin{equation}
\frac{d^5\sigma}{d\Omega_f dE_f d\Omega} = \Gamma (\sigma_T + \epsilon
\sigma_L + v_{LT}\sigma_{LT} \cos\phi^* 
 +  \epsilon \sigma_{TT} \cos 2\phi^* 
 + h p_{e} v_{LT'}\sigma_{LT'} \sin \phi^*)
\end{equation}
where $\epsilon$ is the transverse polarization of the virtual photon,
$v_{LT}=\sqrt{2\epsilon(1+\epsilon)}$, $v_{LT'}=\sqrt{2\epsilon(1-\epsilon)}$,
$\Gamma$ is the virtual photon flux, $\phi^*$  is the pion center of mass
azimuthal angle with respect to the electron scattering plane, $h$ is the electron helicity, and $p_{e}$ is the magnitude of the electron longitudinal polarization. 
The virtual photon differential cross sections
($\sigma_{T},\sigma_{L},\sigma_{LT},\sigma_{TT},\sigma_{LT'}$) 
are all functions of the
center of mass energy $W$, the four momentum transfer squared $Q^2$,
and the pion center of mass polar angle $\theta_{\pi q}^{*}$ (measured
from the momentum transfer direction). They are bilinear combinations of the multipoles \cite{drechsel_tiator}.

\section{Resonant Multipole Fitting}

The current experiments
\cite{beck,blanpied,warren,mertz,kunz,sparveris,joo,frolov,pospischil,bartsch,
  elsner,stave}  do not have
sufficient polarization data to perform a model independent multipole
analysis and must rely upon models for the non-resonant (background) amplitudes. 
The standard procedure to extract the multipoles is to use the models
to fit the data.  Their background terms are
unaltered and the three isospin =3/2 resonance multipoles 
\begin{equation}
R_{i}^{3/2} = M_{1+}^{3/2}, E_{1+}^{3/2}, S_{1+}^{3/2}
\end{equation}
are fit to the data. Specifically we introduced  multiplicative
factors, $\lambda(R_{i})$  for the $I=3/2$ multipoles so that the
phase, and hence unitarity, is preserved. For the charge channels with
a proton target and an outgoing neutral pion   (e.g. $\gamma +p \rightarrow \pi^{0} p$):  
\begin{equation}
R_{i}(\pi^0 p) = R_{i}^{1/2} +\frac{2}{3} \lambda(R_{i}) R_{i}^{3/2}
\end{equation}
where $R_{i}$ represents any of the three photo-pion multipoles $M_{1+}, E_{1+}, S_{1+}$ for the final charge state or in the isospin 1/2 and 3/2 states. 

\begin{figure}
  \begin{tabular}{c}
  \includegraphics[height=.4\textheight]{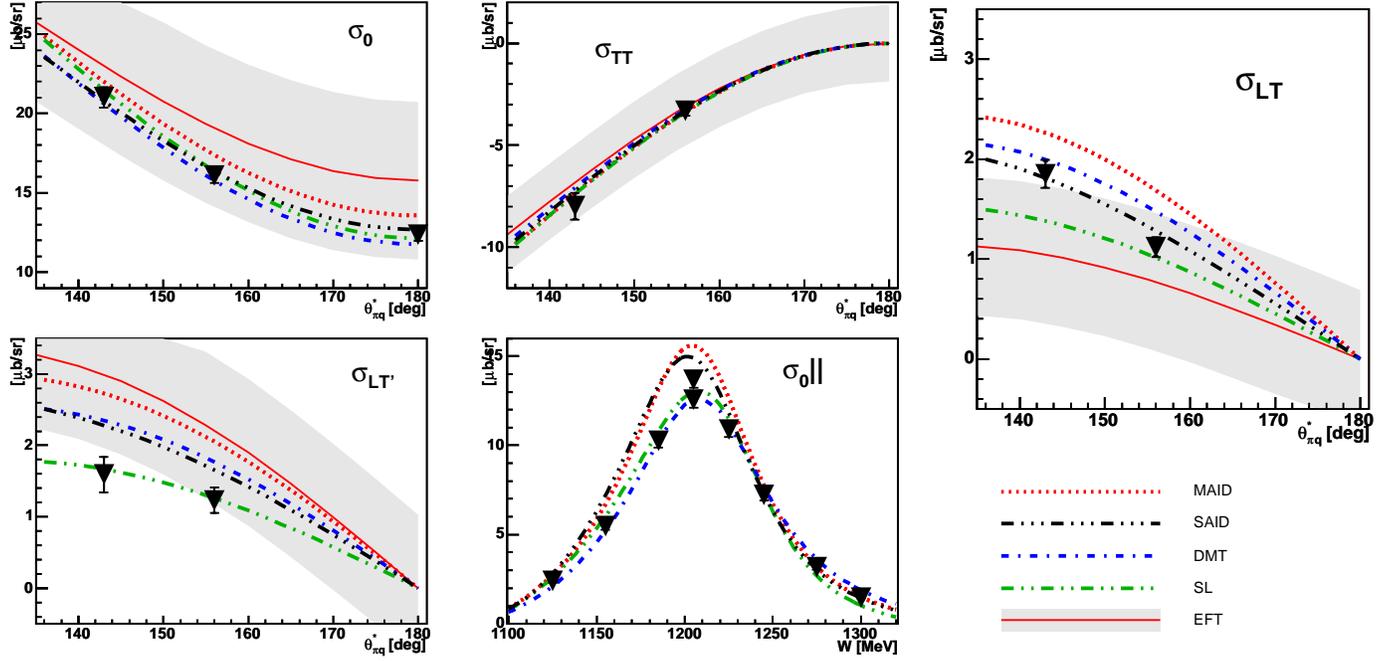}\\
  \hline
  \includegraphics[height=.4\textheight]{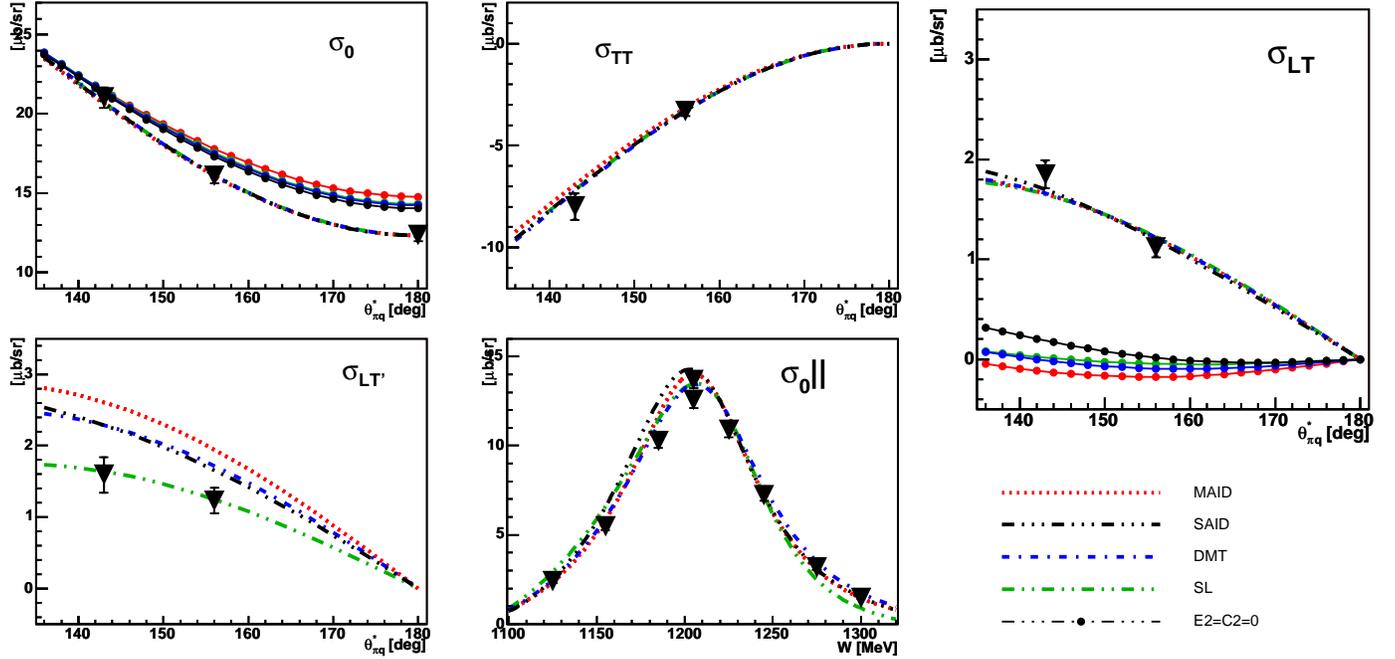}
  \end{tabular}
  \caption{\label{fig:before_after_fit}$Q^2=0.060$ (GeV/c)$^2$ data with
  model predictions before fitting (top panels) and after the three resonant
  parameter fit (bottom panels) along with the EFT predictions from
 Pascalutsa and Venderhaeghen (PV) \cite{pasc}.  
 Note the convergence of the models except
for the background sensitive $\sigma_{LT'}$ points. Data from
\cite{stave} and include the statistical and systematic errors.  The lines with dots on them are the fitted models with
the $E_{1+}$ and $S_{1+}$ quadrupole terms set to zero.  The models
are MAID 2003 \cite{maid1}, SAID \cite{said}, DMT \cite{dmt} and
Sato-Lee (SL) \cite{sato_lee}.
}
\end{figure}

Three parameter, resonant multipole fits were performed on data taken 
at $Q^2=0.060$ \cite{stave}
and $Q^2=0.126$ (GeV/c)$^2$
\cite{warren,mertz,kunz,sparveris,stave_thesis}, one $Q^2$ value at a time,
using four representative calculations: the phenomenological  MAID 2003 \cite{maid1} and SAID\cite{said}models, and the dynamical models of Sato-Lee \cite{sato_lee} and
DMT \cite{dmt}.  The fits are presented in terms of  $M_{1+}^{3/2}$, EMR = E2/M1 =
Re($E_{1+}/M_{1+})$, and CMR = C2/M1 =
Re($S_{1+}/M_{1+})$.  At least one multipole is expressed in
absolute terms rather than as a ratio because some models can give
accurate predictions for ratios but not for absolute sizes. 
Figure \ref{fig:before_after_fit} shows our new, low $Q^2$ data along with
several model predictions before and after the three resonant
parameter fitting.  The convergence is
rather significant.  Only the data points in the top three plots were
included in the fits and yet the parallel cross section as a function
of the center of mass energy $W$ converged nicely.  Note that since
the Sato-Lee model does not include higher resonances, it was not
expected to fit the data well at higher $W$ explaining the deviation
observed in Fig. \ref{fig:before_after_fit}.  Also, as 
expected, the $\sigma_{LT'}$ curves did not converge 
since this time reversal odd observable \cite{rask_don} is primarily sensitive to
background amplitudes and the fit is only for resonant amplitudes.

\begin{figure}
  \includegraphics[height=.4\textheight]{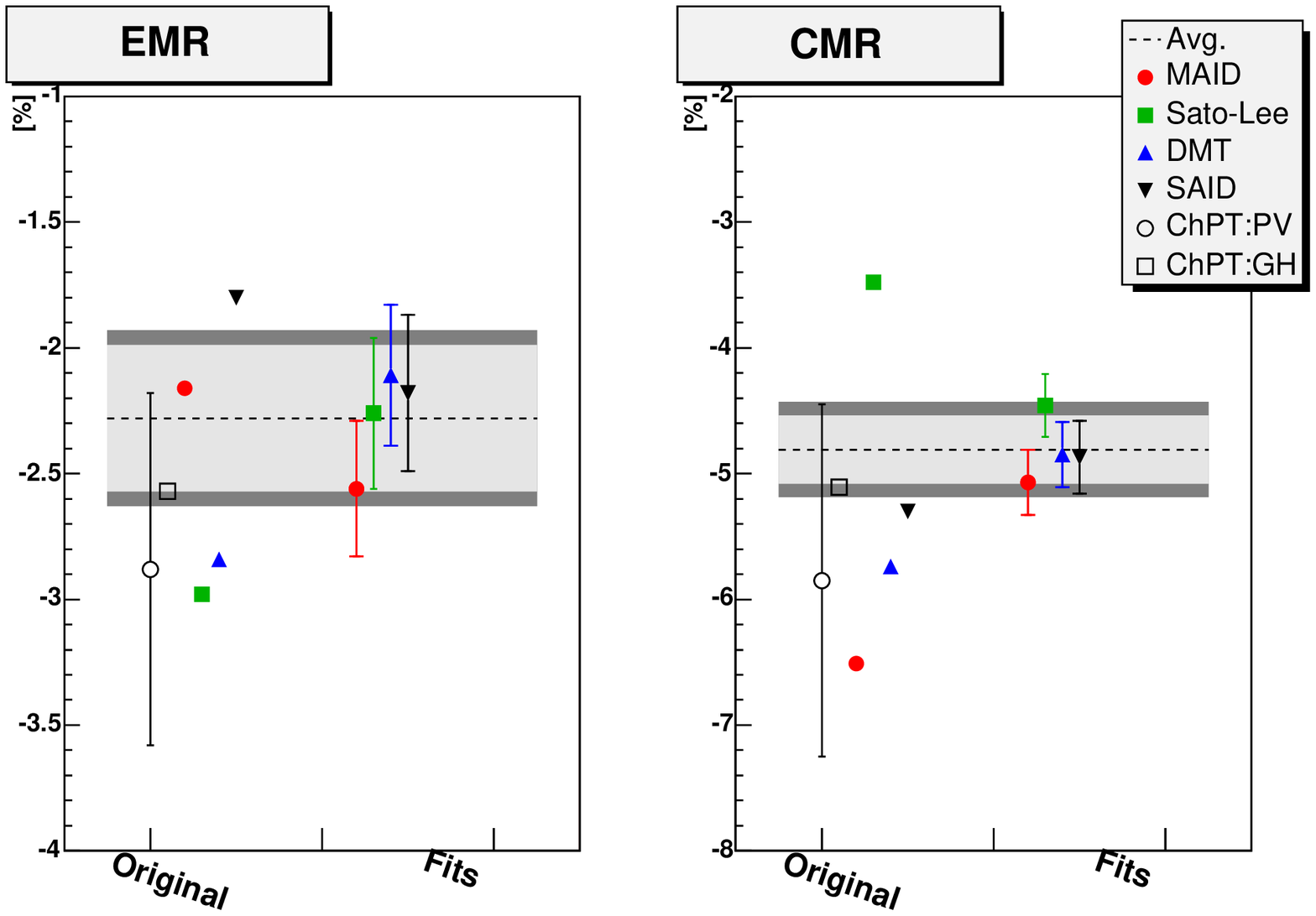}
  \caption{\label{fig:EMR_CMR}$Q^2=0.060$ (GeV/c)$^2$ extracted EMR
  and CMR before and after fitting.  The light error band is the
  average of the fitting errors and the darker band is the RMS
  deviation of the models added in quadrature.   The models
are MAID 2003 \cite{maid1}, SAID \cite{said}, DMT \cite{dmt} and
Sato-Lee \cite{sato_lee} and the effective field theory calculations
of Pascalutsa and Vanderhaeghen (PV) \cite{pasc} and Gail and Hemmert
(GH) \cite{gail_hemmert}.  
The convergence of the models after fitting is clear.}
\end{figure}

The bottom half of Figure \ref{fig:before_after_fit}  shows
the ``spherical'' calculated curves when the  resonant quadrupole
amplitudes ($E_{1+}^{3/2}$ in $\sigma_{0}$ and $S_{1+}^{3/2}$ in
$\sigma_{TL}$, see Appendix for multipole expansions of the
observables) 
are set equal to zero. The difference between the
spherical and full curves shows the sensitivity of these cross sections
to the quadrupole amplitudes and demonstrates the basis of the
 measurement of the $E_{1+}$ and $S_{1+}$ multipoles. The small spread in the spherical curves indicates their sensitivity to the model dependence of the background
amplitudes.
Figure \ref{fig:EMR_CMR} shows the values  for the  EMR
and CMR for the four models before and after fitting.  It is seen that there is a very strong convergence of these values after fitting. We have quoted the average value of these parameters as the measured value and are using the RMS deviation to estimate the model-to-model error\cite{sparveris,stave} since these four models are sufficiently different to have a reasonable estimate of the present state of model dependence of the multipoles. At the present time the model-to-model and experimental errors are approximately equal.    

In a way what we are observing is the fact that the electro-pion
production process shows us two separate faces, depending on the
observable and on the center of  mass energy $W$ that we choose. The
best way to extract the three resonant  amplitudes is to measure the
time reversal even observables ($\sigma_{0}, \sigma_{TT},
\sigma_{LT}$) \cite{rask_don} at or near the resonance energy $W = 1232$ MeV. On the
other hand, the best way to test the model calculations is to examine
time reversal odd observables such as $\sigma_{LT'}$ \cite{rask_don} right on
resonance. In addition, we also have off resonance data. These are
sensitive to both the shape and phase of the $M_{1+}$ multipole and
also the background amplitudes. The $\gamma p \rightarrow \pi^{+}n$
charge channel is also more sensitive to the background amplitudes
particularly the $I = 1/2$ amplitudes. Such background sensitive data in
combination with model studies are essential if the field is to
progress to the stage where the model errors are significantly smaller
than the experimental ones.

\FloatBarrier
\section{Intrinsic Model Errors in Determination of the Resonant Multipoles}

\subsection{Beyond Three Parameter Fits: Including Background Multipoles }
This work expands the three resonant parameter fits to
include the influence of the background multipoles on the resonant amplitudes derived from fitting the experimental data. In this way we will be able to make reasonable estimates of the intrinsic model errors due to uncertainties in the background multipoles and to see if this leads to any suggestions to reduce them. 
First, we include the remaining s and
p wave multipoles:  $E_{0+}, L_{0+}$, $M_{1-},L_{1-}$.  Next, we
estimate the influence of the higher partial waves using the CGLN
invariant amplitudes $F_{i}, i=1,\dots,6$ \cite{drechsel_tiator}.  We 
introduce a new combination of higher order multipoles we call
$\overline{F_{i}}$.  These combinations show that many small multipoles
  can have a cumulative effect as will be seen shortly.  The $\fb{i}$s
  are varied using a scaling factor $\lambda_i$ as
\begin{eqnarray}
 F_i & = & F_i^{S\&P} + \lambda_i\fb{i}  \nonumber  \\
\hline
F_1 & = & \eOp + 3{\mlp}\cos\theta + 3{\elp}\cos\theta + \lambda_1\fb{1} \nonumber \\
F_2 & = & \mlm + 2{\mlp} + \lambda_2\fb{2} \nonumber \\
F_3 & = & 3({\elp} - {\mlp}) + \lambda_3\fb{3} \nonumber \\
F_4 & = & \lambda_4\fb{4} \nonumber \\
F_5 & = & \lOp + 6{\llp}\cos\theta + \lambda_5\fb{5}\nonumber \\
F_6 & = & \llm - 2 {\llp} + \lambda_6\fb{6}.
\end{eqnarray}

Next, we allow the $I=1/2$ part of the charge channel multipole to
vary in a way similar to the $I=3/2$ part.
\begin{equation}
R_i(\pi^0p) = \lambda(R_i^{1/2}) R_i^{1/2} + \lambda(R_i) \frac{2}{3}R_i^{3/2}.
\end{equation}

This new fitting procedure introduces thirteen
 background amplitudes, too many to
determine with the available data.  So, we set out to determine how much
the resonant parameters are affected by the uncertainties in the
background amplitudes.  We try to quantify this effect and 
to determine which parameters have a strong effect and which are
correlated.  

\begin{figure}
\includegraphics[angle=0,scale=0.3]{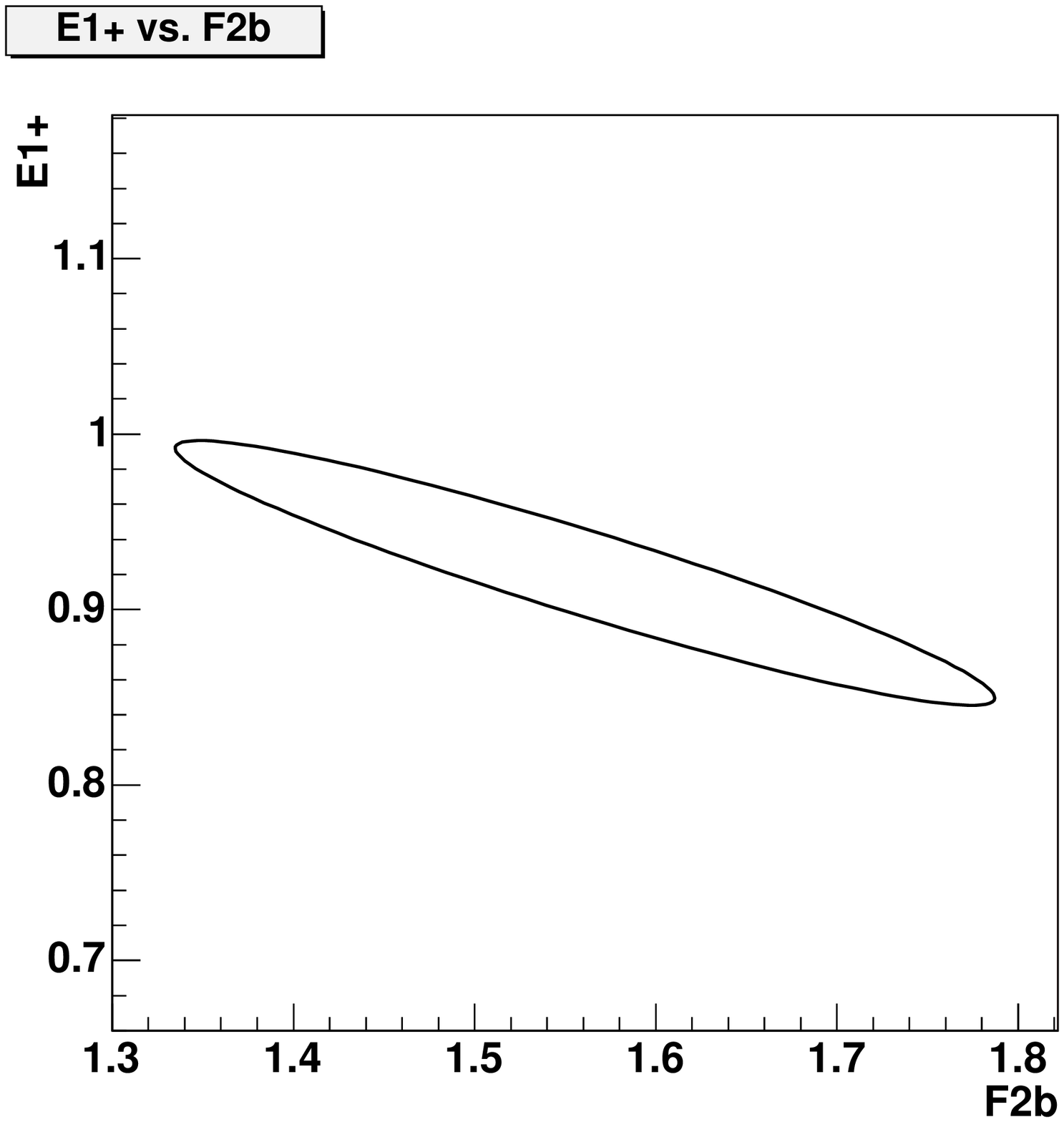}
\includegraphics[angle=0,scale=0.3]{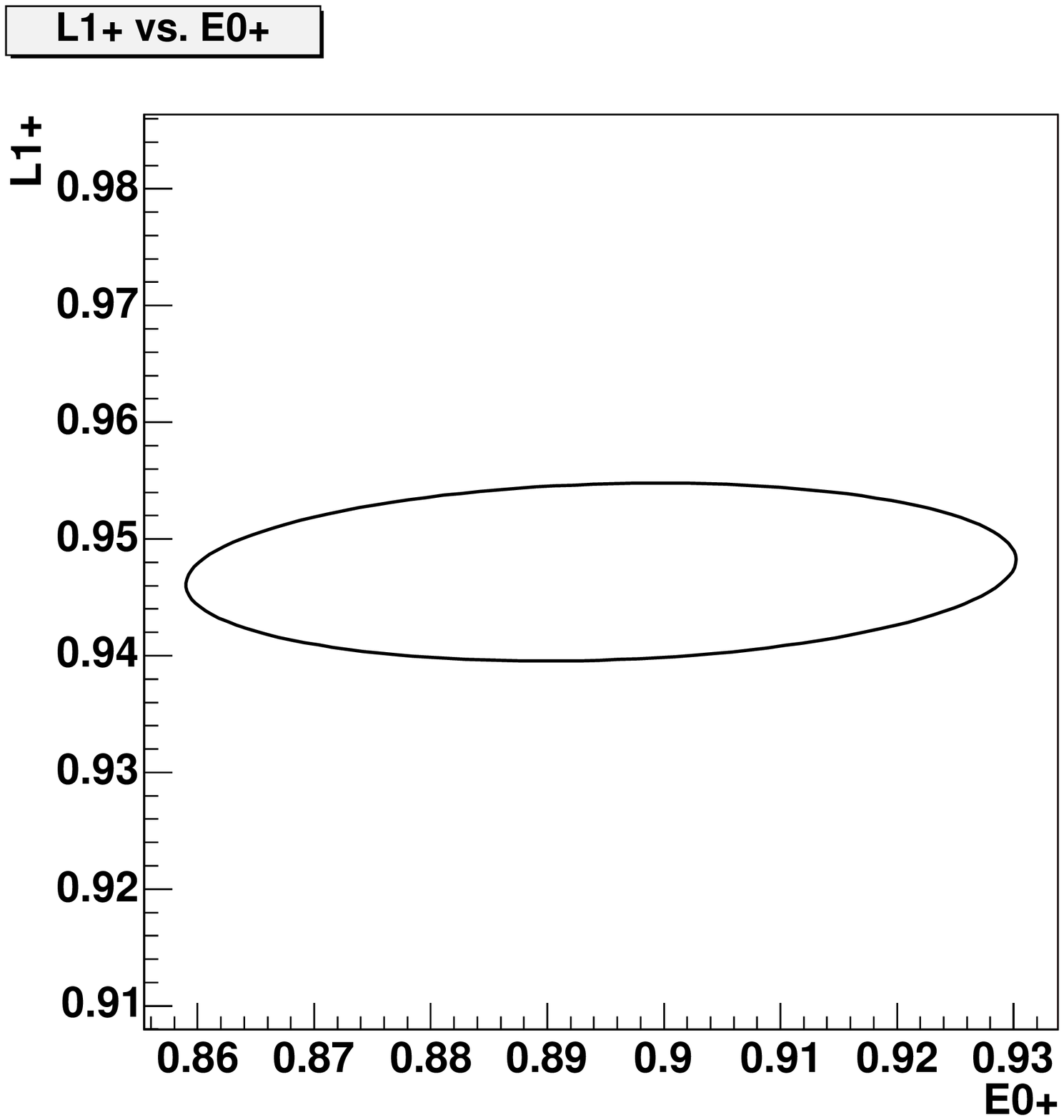}
\includegraphics[angle=0,scale=0.3]{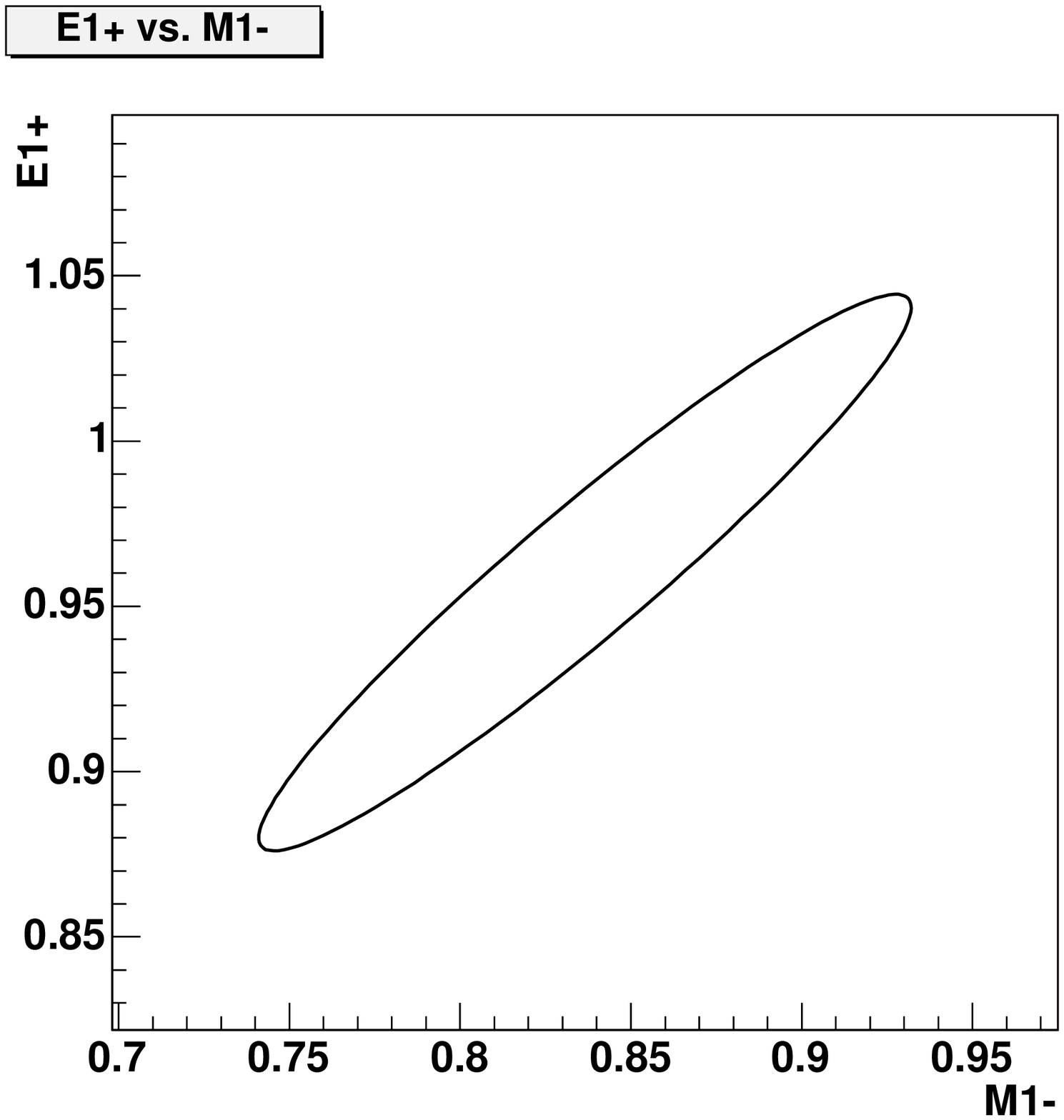}\\
  \caption{\label{fig:corr}Examples of negative correlation (left), no
    correlation (center), and positive correlation (right). Note: F2b=$\fb{2}$. }
\end{figure}

\begin{figure}
\includegraphics[angle=0,scale=0.5]{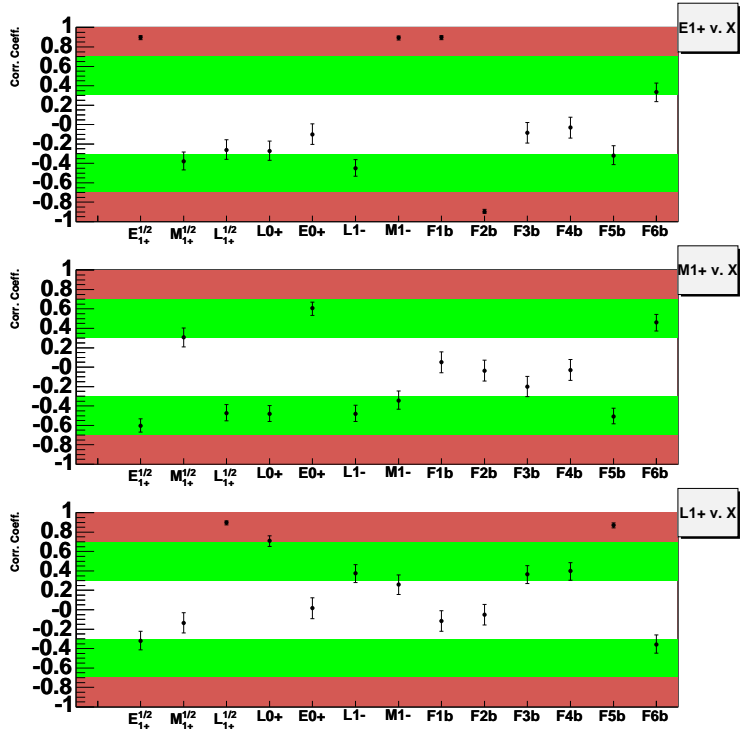}
  \caption{\label{fig:corr_coeff}Correlation coefficients for
    background amplitudes with the resonant amplitudes
    $E_{1+}^{3/2},M_{1+}^{3/2},L_{1+}^{3/2}$ using MAID 2003
    for
    Mainz and Bates $Q^2=0.126$ \gevc data.  The central region of
    each plot is for small correlation with the next region being
    medium correlation and the next being large correlation. Note:
    F1b=$\fb{1}$, F2b=$\fb{2}$, etc.
}
\end{figure}

First, we look for parameters which are correlated with the resonant
parameters.  Figure \ref{fig:corr} shows some examples of negative,
positive and no correlation.  Fitting parameters are plotted on each
axis and the ellipse indicates the region where 68\% (1 $\sigma$) of
the fits are expected to fall if many similar data sets are fit.  The
ellipse with axes close to the x and y axes shows no correlation.
However, the other ellipses are rotated indicating that as one
parameter tends one direction, the other parameter tends to go with it
or away from it.  This indicates a correlation between the parameters.
While error ellipse plots are useful in
a qualitative way, they are difficult to use in a
quantitative manner.  Changes in the scale of the axes will change the
angle of the ellipse hiding or exaggerating correlations.  
Therefore, we use the correlation
coefficient, $r$, to indicate the level of correlation between two parameters:
\begin{equation}
r=\frac{\sigma_{12}}{\sigma_{11}\sigma_{22}}
\end{equation}
where the error and curvature matrices 
\begin{equation}
\left(\begin{array}{cc}
\epsilon_{11} & \epsilon_{12} \\
\epsilon_{21} & \epsilon_{22}
\end{array}\right)^{-1} 
= 
\left(\begin{array}{cc}
\sigma_{11}^2 & \sigma_{12} \\
\sigma_{21} & \sigma_{22}^2
\end{array}\right)
\end{equation}
are used \cite{PDH}.  $r$ varies from -1
to 1 and is insensitive to the parameter scales since the scale factor
for each parameter
cancels in the ratio.  Figure \ref{fig:corr_coeff} shows the various
correlation coefficients for each background amplitude with $E_{1+}^{3/2},
M_{1+}^{3/2}$, and $L_{1+}^{3/2}$ using MAID 2003 and combined Bates and Mainz 
data at $Q^2=0.126$ \gevcp.  
The square of $r$  indicates how much of the
variance is explained by a linear relation between the two variables.
A rule of thumb is that two variables are correlated if $r^2 \ge
0.5$ and uncorrelated if $r^2 \le 0.1$.  This then
leads to the ranges in $r$: 
$0.7 \le |r| \le 1.0 =$ large correlation,
$0.3 \le |r| < 0.7 =$ medium correlation, 
$0.0 \le |r| < 0.3 =$ small correlation.

\begin{figure}
\includegraphics[angle=0,scale=0.65]{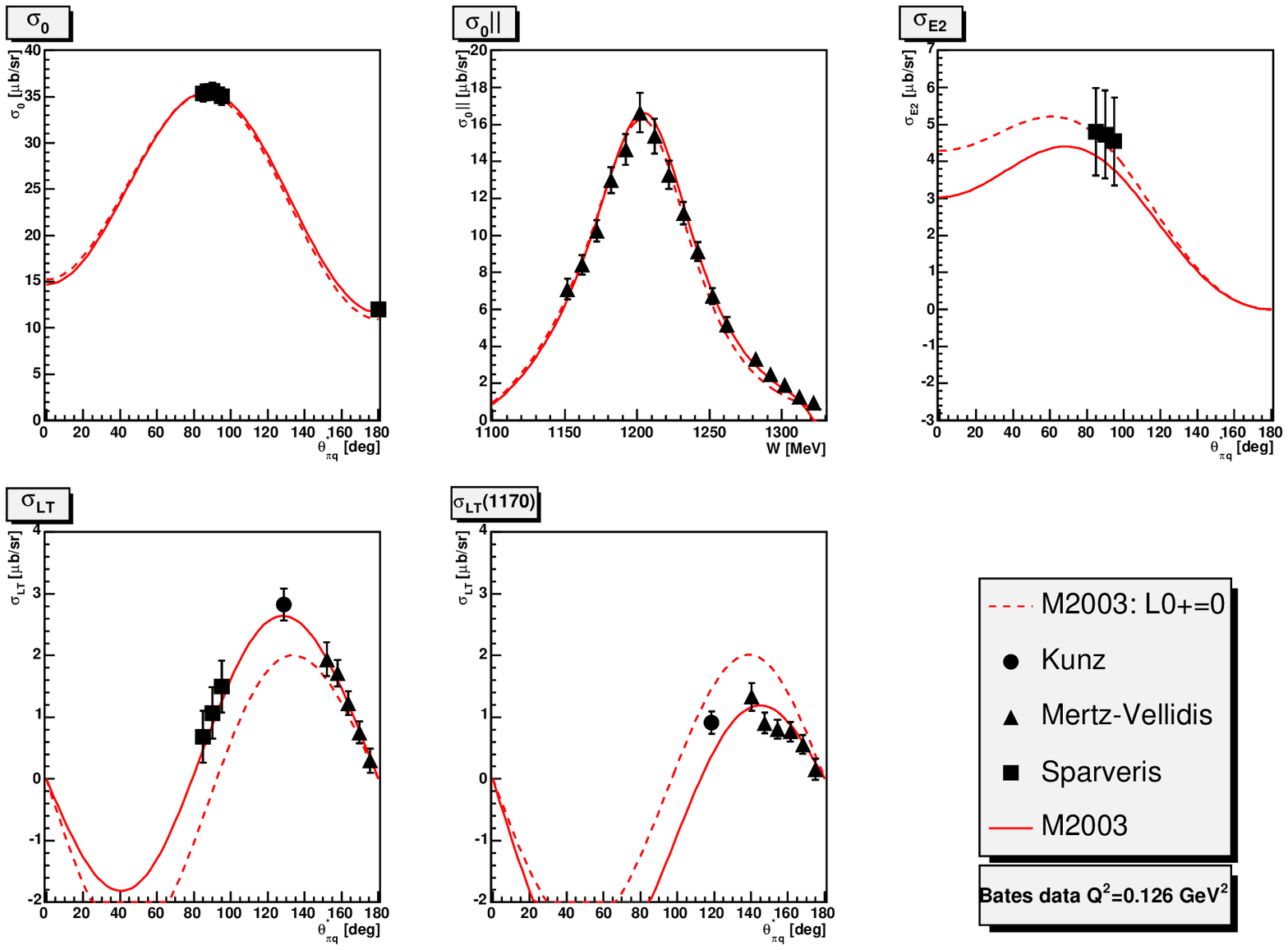}
  \caption{\label{fig:sens_L0}Sensitivity to $L_{0+}$ in MAID 2003 
using data at
    $Q^2=0.126$ \gevcp.  The solid curve is the full MAID 2003 model
    and the dashed line is with the $L_{0+}$ multipole set to
    zero. The effect is particularly evident for $\sigma_{LT}$.  Data
    are from \cite{mertz,kunz,sparveris} and include statistical,
  systematic and model errors.
}
\end{figure}

\begin{figure}
\includegraphics[angle=0,scale=0.65]{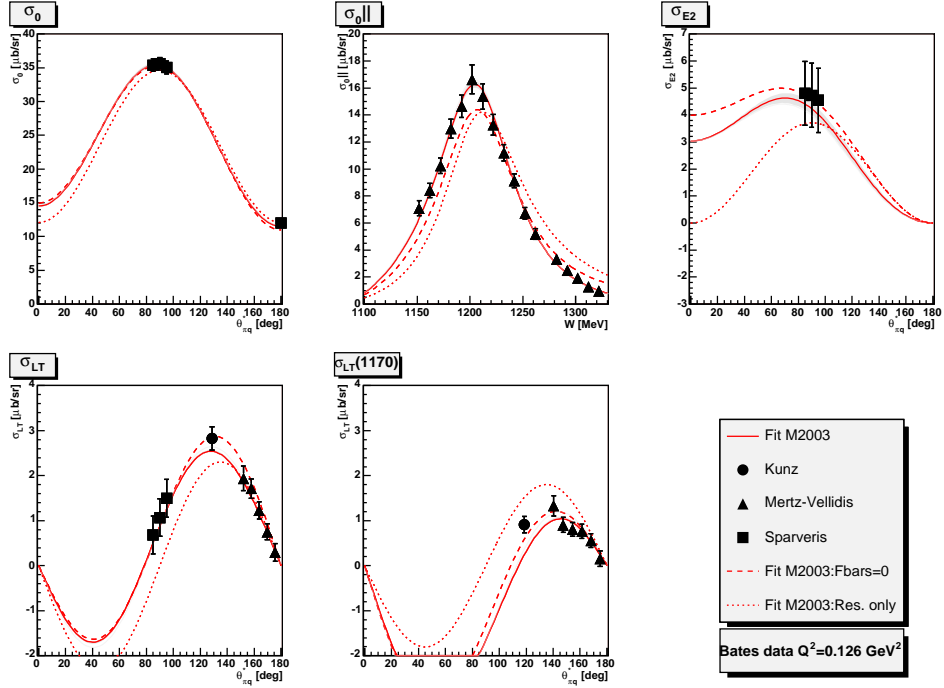}
  \caption{\label{fig:sens_other}Sensitivity to $\fb{i}$ and s and p
    in MAID 2003 using data at
    $Q^2=0.126$ \gevcp.  As in Fig. \ref{fig:sens_L0}, the solid curve
  is the full MAID 2003 while the dashed line is now MAID 2003 with
  all the $\fb{i}$s turned off and the dotted line shows the results
  using only the resonant multipoles.  This shows the significant
  effect the background has on some observables and that the $\fb{i}$
  amplitudes can also have effects of similar size.  Data
    are from \cite{mertz,kunz,sparveris} and include statistical,
  systematic and model errors.
}
\end{figure}

The next check is for sensitivity.  If a parameter is large, zeroing
it out should affect the $\chi^2$ by a large amount.  For example,
when $L_{0+}$ is turned off, the model predictions
change noticeably (see Fig. \ref{fig:sens_L0}).  
Other background terms can have significant
effects as well like the $\fb{i}$s and the remaining s and p do in
Fig. \ref{fig:sens_other}.  Table \ref{table:sens} shows the
$\chi^2$/d.o.f. that results from turning off the various background
amplitudes in the MAID 2003 model.  
It also indicates how strongly the amplitude was
correlated with any of the resonant multipoles.

In Figs. \ref {fig:sens_L0} and \ref{fig:sens_other}, a combined cross
section\cite{cnp,mertz,sparveris}  $\sigma_{E2}=\sigma_0(\theta_{\pi
  q}^*) + \sigma_{TT}(\theta_{\pi q}^*) -
\sigma_0(\theta_{\pi q}^*=\pi)$ is shown. In this linear combination the dominant $M_{1+}$
multipole contribution cancels out and shows the effect of the smaller $E_{1+}$ quadrupole contribution.  (See Appendix for the expansion of the observables
in terms of multipoles.)

\begin{table}
\begin{tabular}{cc}
Extra Par. & $\chi^2$/d.o.f\\
\hline
$L_{0+}$ & {\bf	7.42}\\
$E_{0+}$ &	6.09\\
$\fb{1}$ &	{\bf 5.60}\\
$L_{1+}^{1/2}$ & {\bf	5.59}\\
$M_{1-}$ & 	{\bf 4.10}\\
$E_{1+}^{1/2}$ & {\bf	3.31}\\
$\fb{5}$ &   {\bf 1.85} \\
$M_{1+}^{1/2}$ & 	 1.85 \\
$\fb{2}$ &	{\bf 1.80} \\
$\fb{6}$ & 1.72 \\
$L_{1-}$ & 	 1.55 \\
$\fb{3}$ &	1.24\\
--- & 1.21\\
$\fb{4}$ &   1.17 \\
\end{tabular}
\caption{\label{table:sens}Gives $\chi^2/$d.o.f. resulting from
turning off the corresponding background parameter in MAID 2003 arranged from
most to least sensitive using the
combined Bates and Mainz $Q^2=0.126$ \gevc data.  The type face indicates the level
of correlation with any of the three resonant multipoles:
{\bf Large Correlation} Medium or Small Correlation.
}
\end{table}

Using the criteria of correlation ($|r|>0.7$) and sensitivity
($\chi^2$ increases by 50\% upon removal), several parameters were identified as
significant using the $Q^2=0.126$ (GeV/c)$^2$ data set.  Those
multipoles are shown in Table \ref{table:sens_filter} and include two of the
s and p multipoles, two of the isospin = 1/2 multipoles and three
$\fb{i}$ terms.  Looking at Fig. \ref{fig:res_pars} all seven terms
shown in Table \ref{table:sens_filter}, when varied, lead to shifted central
values or larger error bars for the resonant multipoles.  Some, like
the $E_{1+}^{1/2}$ with $M_{1+}^{3/2}$ are shifted but not outside the error
bars and with no increase in the error bar size and so are not
considered significant.

Effort was made to search for a set of criteria using the correlation
coefficient and the change in $\chi^2$/d.o.f. that would identify all
of the parameters which Fig. \ref{fig:res_pars} identifies as
significant.  The criteria for significance were an increased error
and/or a shift in central value.  Either indicates a significant
effect on the resonant multipole determination.  
In order to make the criteria robust, other models were
put through the selection process as well.  In addition to the MAID
2003 model, Sato-Lee, SAID, and DMT were all used.  The best
identifier of significant parameters turned out to be the single test
of $|r|>0.7$.  In almost every case, this alone identified all the
significant parameters.  The $\chi^2$ sensitivity would identify some
of the sensitive parameters but not others.  

\begin{center}
\begin{table}
\begin{tabular}{ccc}
\hline
E1+ & vs. & $E_{1+}^{1/2}$, $M_{1-}$, $\fb{1}$, $\fb{2}$\\
L1+ & vs. & $L_{1+}^{1/2}$, $L_{0+}$, $\fb{5}$\\
\hline
\end{tabular}
\caption{\label{table:sens_filter} Background amplitudes for which the quadrupole amplitudes show sensitivity  for MAID 2003
  using the
  criteria listed in the text and the combined Bates and Mainz $Q^2=0.126$ \gevc data.}
\end{table}
\end{center}

\FloatBarrier
\section{Effect of Background on Resonant Amplitudes}
\begin{figure}
\includegraphics[angle=0,scale=0.4]{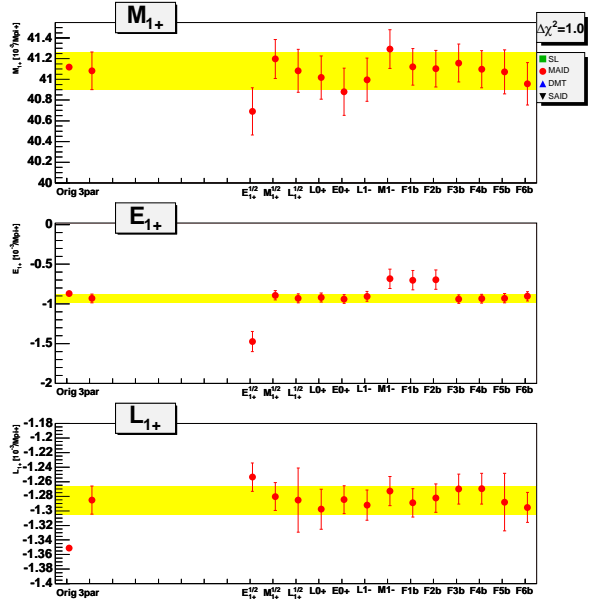}
  \caption{\label{fig:res_pars}Effect of sensitive parameters on
  extracted resonance values using MAID 2003 for the combined Mainz
  and Bates data at
  $Q^2=0.126$ (GeV/c)$^2$.  Note the larger error
  bars and shifted central positions of certain combinations of
  multipoles.  The horizontal band indicates the value and error
  resulting from the three resonant parameter fit.  Similar plots were
  made for the remaining models.  Their results are shown in
  Fig. \ref{fig:res_par_all} and Table \ref{table:res_par_all}.
}
\end{figure}

In order to see the effect the sensitive background amplitudes have on the
extracted results, the resonant parameters resulting from each four
parameter fit were plotted in
Fig. \ref{fig:res_pars}.  The horizontal bar indicates the position and
error of the three parameter fit.  The 
background amplitudes identified as significant do have an effect on
the extracted multipoles relative to the three parameter fit.  For
each sensitive background parameter, the error increases and in most
cases the central values shift.  What is also interesting is that the
$\fb{i}$s have a significant effect.  This indicates that many small
amplitudes can combine to have a large effect.

\begin{figure}
\includegraphics[angle=0,scale=0.4]{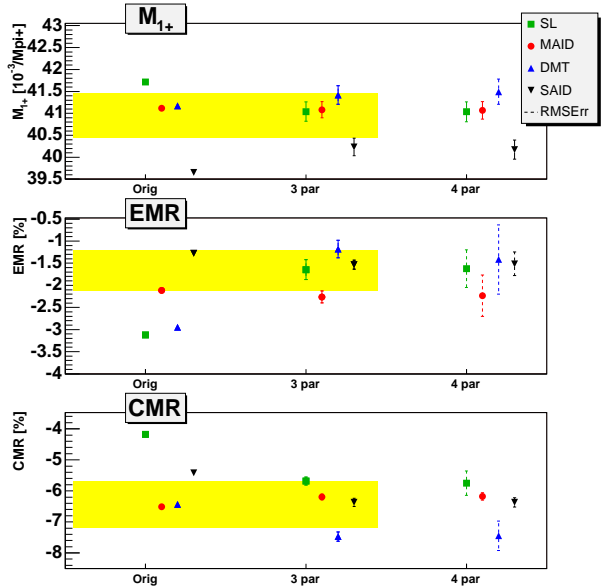}
  \caption{\label{fig:res_par_all}Comparison of the estimated
  intrinsic and model-to-model errors for all four models at
  $Q^2=0.126$ (GeV/c)$^2$, $W=1232$ MeV.  The
  intrinsic error was estimated using the RMS deviation across all the
  four parameter fits.  If the deviation was small, then the average
  of the four parameter fit errors was used.  Cases where the RMS
  deviation was used are indicated by the dashed vertical error
  lines.  The horizontal band indicates the RMS deviation of the three
resonant parameter fits using the four different models.}
\end{figure}

To try to quantify the effect of the various background amplitudes on
the resonant amplitudes, the RMS deviation of the various four
parameter fits was
taken for each model and identified as the intrinsic model error.  For
some fits, the RMS deviation was small and so the average of the four
parameter fitting errors was used instead.  In both cases, an estimate of
the intrinsic error in each model was obtained.
The results are shown in
Fig. \ref{fig:res_par_all} and Table \ref{table:res_par_all} 
along with the average and RMS deviation of
the three parameter fits (model-to-model error).  
The figure and table indicate that the  model-to-model variation is about the same size as the intrinsic model error (specifically for  $M_{1+}$ and the  CMR the model-to-model error is larger while for the EMR it is somewhat smaller).  However, the new
error determination procedure is able to use one model alone instead of
comparing it with other models.  Each models's error can be assessed
independently of the other models.

\begin{center}
\begin{table}
\begin{tabular}{clcc|cccc}
         &                   & 3 par. avg. & Model-to-model & Intrinsic errors \\
         &                   &             &  error   &  MAID & DMT & Sato-Lee & SAID\\
\hline
$M_{1+}$ & [$10^{-3}/m_{\pi^+}$] & 40.94 & 0.50 &  0.20 & 0.29 & 0.23 & 0.22 \\
EMR &[\%]                    & -1.65 & 0.45 &  0.47 & 0.78 & 0.43 & 0.27 \\
CMR &[\%]                    & -6.43 & 0.75 &  0.12 & 0.47 & 0.39 & 0.15 \\
\hline
\end{tabular}
\caption{\label{table:res_par_all}Preliminary fitting results and 
comparison of the estimated
  intrinsic and model-to-model errors for all four models for the
  combined Mainz and Bates data
  at $Q^2=0.126$ (GeV/c)$^2$, $W=1232$ MeV.  See
  Fig. \ref{fig:res_par_all} for calculation details.  
}
\end{table}
\end{center}

While looking to improve the fits, an exhausive search was performed of
all combinations of the three resonant parameters and any combination
of the 13 remaining parameters.  No significant improvement was found
for either $Q^2=0.060$ or $0.126$ (GeV/c)$^2$.  

It is time, then, to look beyond fitting the multipoles.  It is
possible to modify internal model parameters (form factors, coupling
constants) which affect many multipoles simultaneously but in
different ways.  This may allow the models to fit the data better.
However, this fitting most likely needs to be performed by the model authors.

The understanding of the $\Delta$ will also be improved with experiments that are
closer to complete.  With target and recoil polarization, more
observables are accessable and these have different combinations of
multipoles.  These new combinations will further constrain the models
allowing better fits and smaller uncertainties in the backgrounds.
Until new data are available, though, fitting the data and improving
the models remain the best options.

\FloatBarrier
\section{Summary and Conclusion}

Experimental results using the $\gamma^* p\rightarrow p \pi^0$ reaction have advanced the
understanding of the shape of the proton and the $\Delta$.  
However, the analysis process begins with extracting multipoles (which
are not observables) from cross
sections (which are).  Without complete experiments including target and recoil
polarization, the extraction must rely upon models for the background
amplitudes.  Performing standard
three resonant parameter fits has allowed a good deal of progress to
be made.  Near resonance, fits using various models converge at
$Q^2=0.060$ and 0.126 (GeV/c)$^2$ despite the differences in the model backgrounds.  
However, what has not been fully understood is the effect these
differing backgrounds can have on the resonant parameters.

To answer that question, we have added thirteen more background
amplitudes  to our three parameter fits and systematically examined
the effect of each one on all three resonant multipoles.  Those
additional background amplitudes
are the four remaining s and p wave amplitudes, three isospin 1/2
amplitudes and six amplitudes we have constructed, the $\fb{i}$s.  The
large effect of some of the $\fb{i}$ terms shows how small
multipoles which may have been ignored separately, can combine to have
a sizable effect on the resonant amplitudes.

As part of the systematic examination of the additional background
amplitudes, correlations were found between them and the resonant
amplitudes which led to larger errors and/or shifts in the values of
the extracted resonant multipoles.  We also found that while some
amplitudes exhibit a large sensitivity in $\chi^2$, no universal criteria
could be found which would predict a sensitivity in the resonant
amplitudes.  Some background amplitudes which were sensitive did not
affect the fits while others which were not sensitive did.

However, varying the background amplitudes which were highly
correlated with the resonant multipoles did affect the
extracted resonant multipoles.  Previous works have shown
 that the experimental and model-to-model errors
are similar in size \cite{stave,sparveris}.  
What this exercise has shown is that the
intrinsic model error is also similar in size to the model-to-model
error.  The current data really are challenging the existing models.  
So, without improvement in the models or more complete
experiments, this is as far as the current data can take us.

In general, the models agree with the data in a qualitative way but a good
quantitative decription will require further refinement of those
models.  It is possible that some of the models may be made
to fit the data much better with adjustment of the proper parameters.  Adjusting a
form factor or coupling constant within the model will change many
multipoles in ways that are different from how they were varied in
this study.  
Once the models are improved, they can be further tested with
experiments that utilize target and recoil polarization.  These
introduce new combinations of multipoles which will further constrain
the models.  However, until new data are available, improvement of the
models is the only option which will allow a better understanding of
the $\Delta$.

\begin{figure}
\includegraphics[angle=0,scale=0.4]{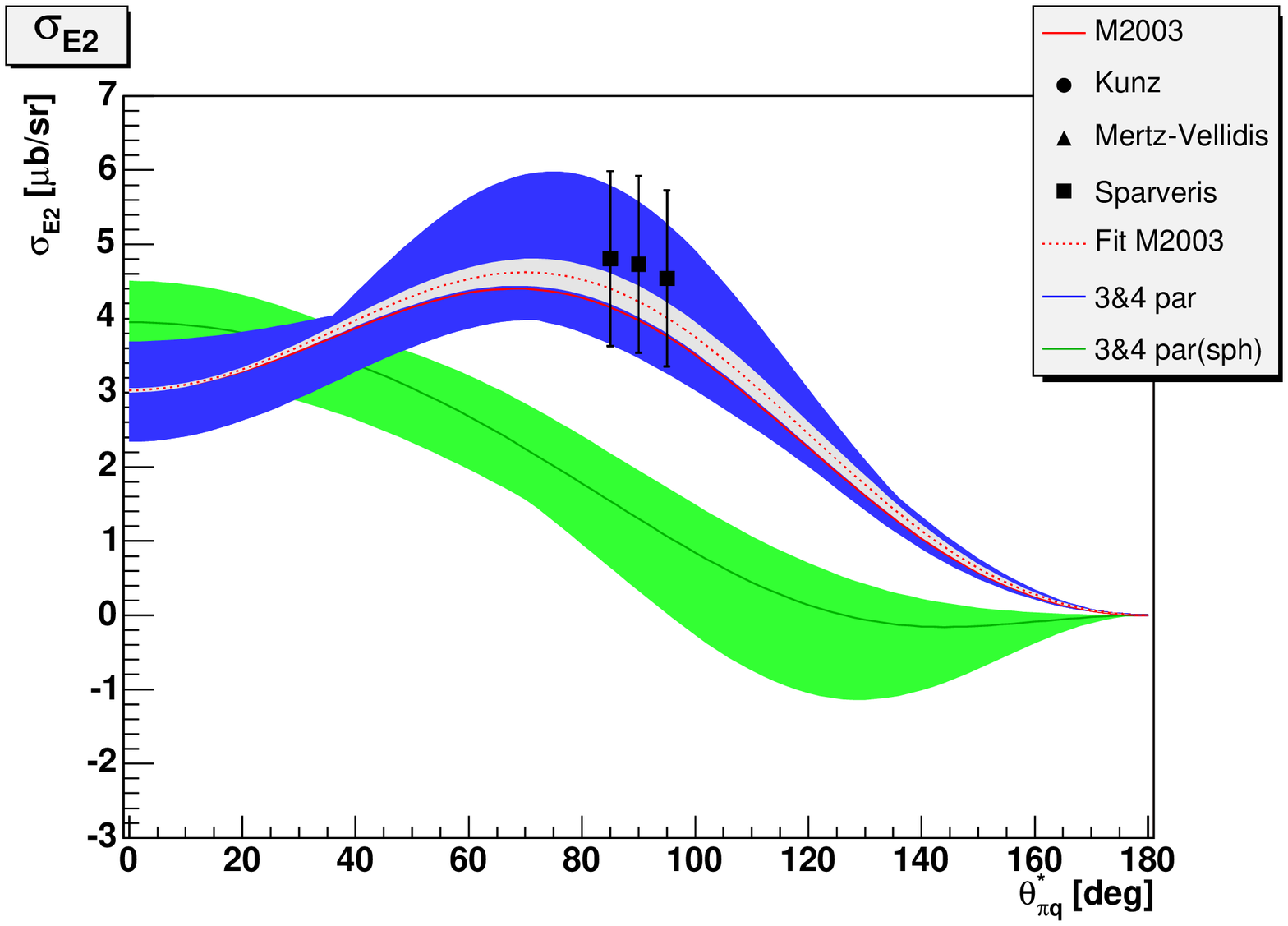}
\includegraphics[angle=0,scale=0.4]{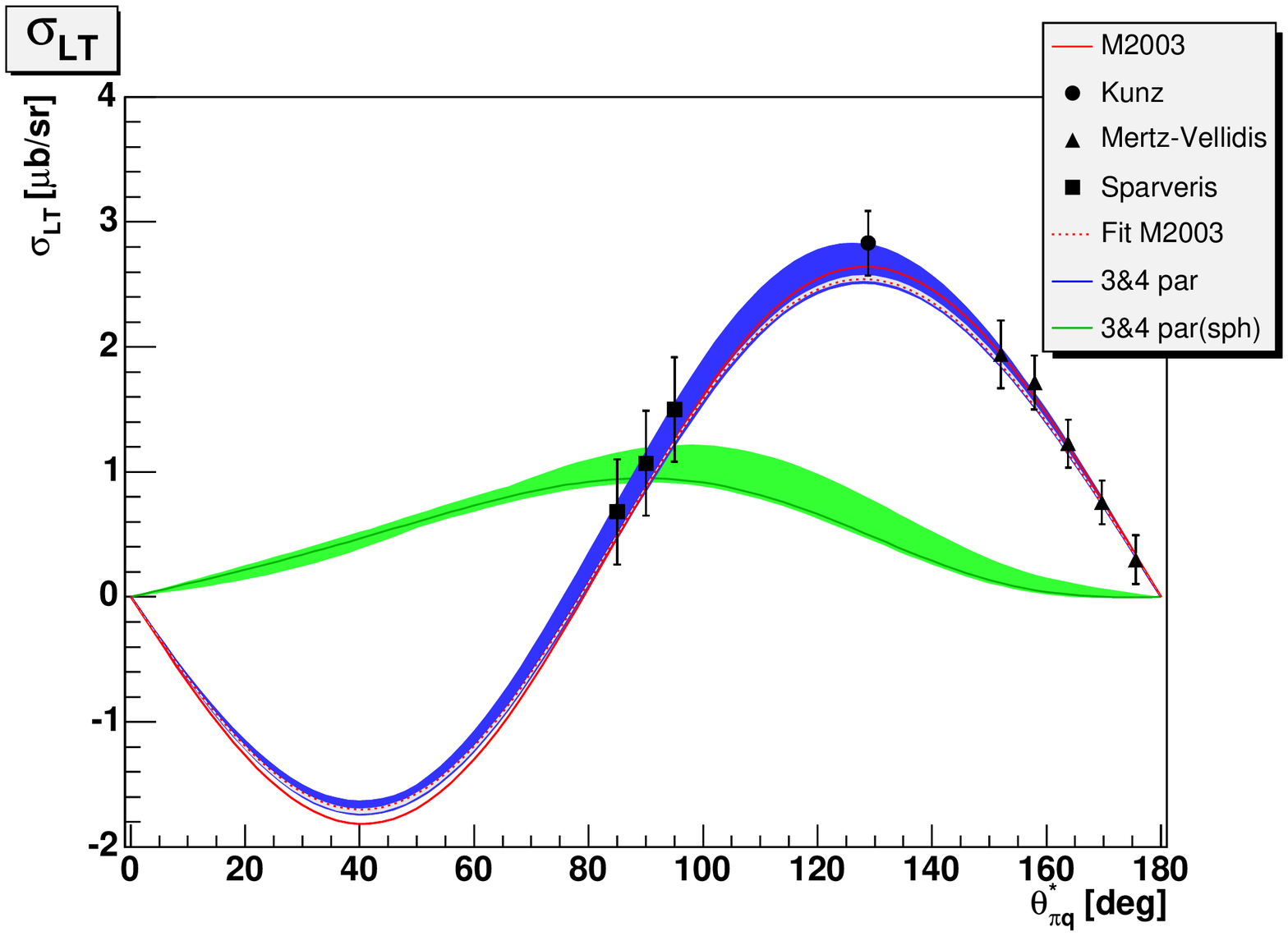}
  \caption{\label{fig:M2003_bands}$\sigma_{E2}$ (left) and
  $\sigma_{LT}$ (right) with error bands showing the range of
  all the MAID 2003 four parameter fits with (blue/dark region) and
  without (green/gray region) the
  quadrupole terms.  The data are clearly in the region indicating
  quadrupole strength similar in size to that found in the models.
  These plots represent fits using the MAID 2003 model.  The light
  region surrounding the MAID 2003 fit was found using the errors in
  the three parameter fit with fixed background terms.
  Data are from \cite{mertz,kunz,sparveris} and include statistical,
  systematic and model errors.
  }
\end{figure}

Finally we return to the question that has primarily motivated this
field: Do we have definitive evidence that the nucleon and $\Delta$ have non-spherical components and if so how large?  Based on this study we follow reference \cite{cnp} and present Figure \ref{fig:M2003_bands} which indicates the final sensitivity to the
quadrupole amplitudes.  On the right is $\sigma_{LT}$ which is
sensitive to the $S_{1+}$ quadrupole term.  On the left is a special
construction, $\sigma_{E2}=\sigma_0(\theta_{\pi q}^*) +
\sigma_{TT}(\theta_{\pi q}^*) -
\sigma_0(\theta_{\pi q}^*=\pi)$\cite{cnp,mertz,sparveris} which cancels out the dominant $M_{1+}$
multipole contribution and shows the effect of the smaller $E_{1+}$ quadrupole
contribution. 
In Fig. \ref{fig:M2003_bands}, the range of predictions using all the four
parameter fits was found by cycling through all the fits and storing
the maximum and minimum.  In this way, a high probability region was
identified where the physical multipole would be expected to be.  For
comparison, the same procedure was repeated but with the quadrupole
amplitudes set to zero. On the basis of this study of the uncertainties in the resonance and background amplitudes we agree with the previous conclusion\cite{cnp} that a significant contribution of quadrupole amplitudes has been observed. 

\begin{theacknowledgments} The authors would like to thank  L. Tiator,
D. Drechsel, T.-S. H. Lee, V. Pascalutsa, M. Vanderhaeghen, T. Gail
and T. Hemmert for their assistance with valuable discussions and for
sharing their unpublished work.   This work is supported  at MIT by
the  U.S. DOE under Grant No. DE-FG02-94ER40818.
\end{theacknowledgments}







\section{Appendix:Contribution of Higher Partial Waves in the  Leading Multipole Approximation}

The response functions can be expanded keeping only the terms which
interfere with the dominant $M_{1+}$ multipole.  The multipoles for
$L\ge2$ have been combined into the $\fb{i}$s in the following
expansions which are called the Leading Multipole Approximation (LMA):

\begin{eqnarray}
R_{T}^{LMA} & = & \left(\frac{5}{2}-\frac{3}{2}\cos^2\theta\right)\abs{\mlp} \nonumber \\
            & + & 6\cos^2\theta{\rm Re}\left[\conj{\elp}\mlp\right]    \nonumber \\
            & - & \sin^2\theta\left(3{\rm Re}\left[\conj{\elp}\mlp\right] +
            {\rm Re}\left[\conj{\mlp}\fb{3}\right]\right) \nonumber\\
            & + & (1-3\cos^2\theta)\left({\rm Re}\left[\conj{\mlp}\mlm\right] +
	    {\rm Re}\left[\conj{\mlp}\fb{2}\right]\right)\nonumber\\
            & + & 2\cos\theta\left({\rm Re}\left[\conj{\eOp}\mlp\right]+
	    {\rm Re}\left[\conj{\mlp}\fb{1}\right]\right)
\end{eqnarray}

\begin{eqnarray}
R_{TT}^{LMA} & = &
-{\rm Re}\left[\conj{\mlp}(3\elp+3\fb{2}+\fb{3}+3\mlm)\right]\sin^2\theta \nonumber \\
             & - & \frac{3}{2}\abs{\mlp}\sin^2\theta
\end{eqnarray}

\begin{eqnarray}
R_{LT}^{LMA} & = &
\sin\theta{\rm Re}\left[\conj{\mlp}(\fb{5}+\lOp+6\llp\cos\theta)\right]
\end{eqnarray}

\begin{eqnarray}
R_{LT'}^{LMA} & = &
\sin\theta{\rm Im}\left[\conj{\mlp}(\fb{5}+\lOp+6\llp\cos\theta)\right]
\end{eqnarray}

\begin{eqnarray}
\sigma_{E2}^{LMA} & = & 
-12\sin^2\theta {\rm Re}\left[\conj{\elp}\mlp\right]\nonumber \\
     & - & 2{\rm Re}\left[\conj{\mlp}\fb{3}\right]\sin^2\theta \nonumber \\
     & + & \left(2\cos\theta
+2\right)\left({\rm Re}\left[\conj{\eOp}\mlp\right] +
{\rm Re}\left[\conj{\mlp}\fb{1}\right]\right)
\end{eqnarray}

\end{document}